\documentstyle[prb,twocolumn,aps,epsf]{revtex}
\begin{document}

\twocolumn[
\hsize\textwidth\columnwidth\hsize\csname@twocolumnfalse\endcsname
\title{Calculated temperature-dependent resistance in low density 2D hole
gases in GaAs heterostructure}
\author{S. \ Das Sarma and E. H.\ Hwang}
\address{Department of Physics, University of Maryland, College Park,
Maryland  20742-4111 } 
\date{\today}
\maketitle

\begin{abstract}
We calculate the low temperature resistivity in low density 2D hole 
gases in GaAs heterostructures by including screened charged impurity 
and phonon scattering in the theory. Our calculated resistance, which 
shows striking temperature dependent non-monotonicity arising from the 
competition among screening, nondegeneracy, and phonon effects, is in 
excellent agreement with recent experimental data. 


\end{abstract}
\vspace{0.5in}
]


A number of recent density-dependent low temperature transport 
measurements in dilute two dimensional (2D) n-Si MOSFET and p-GaAs 
heterostructure systems have attracted a great deal of attention \cite{one} 
because the experiments nominally exhibit a metal-insulator-transition (2D MIT)
as a function of 2D carrier density ($n$). In addition to this unexpected
2D MIT phenomenon (at this stage it is unclear whether the transition 
represents a true $T=0$ quantum phase transition (QPT) or a finite temperature 
crossover behavior) these measurements reveal a number of intriguing 
transport properties \cite{one} in dilute 2D systems, such as a 
remarkable temperature dependence of the low density 
resistivity in the nominally metallic phase, which deserve serious 
theoretical attention in their own rights irrespective of whether the 
2D MIT phenomenon is a true QPT or not.

In this paper we provide a quantitative theory for one such recent 
experiment \cite{two} carried out in a low density GaAs-based 2D hole gas. 
In our opinion, Ref. \onlinecite{two} represents a particularly important 
experiment in relation to the 2D MIT phenomenon (although ironically no MIT 
is actually observed in Ref. \onlinecite{two} --- even the lowest density 
data in Ref. \onlinecite{two} are entirely in the nominally metallic phase) 
because the ultra-pure  samples used in Ref. \onlinecite{two} explore the 
2D ``metallic'' regime of the highest mobility (i.e., the best quality or 
equivalently the lowest disorder), the lowest carrier density, and the 
lowest temperature so far studied in the context of the 2D MIT phenomenon. 
More specifically, there have been suggestions and speculations \cite{one} 
that the 2D MIT phenomenon is an interaction-driven QPT (the scaling theory 
of localization \cite{three} rules out a true localization transition in 
2D disordered system) with the dimensionless $r_s$ 
parameter, which is the ratio of the interaction energy to the 
noninteracting kinetic energy of the 2D electron system, being the tuning 
parameter which drives the QPT. It is important to emphasize that $r_s$ 
increases as $n$ decreases ($r_s \propto n^{-1/2}$), and therefore the 2D 
systems of Ref. \onlinecite{two} represent the highest (lowest) $r_s$ 
($n$) and consequently the most strongly interacting 2D systems 
experimentally studied so far in the context of the 2D MIT phenomenon. 
To be precise, $r_s$ values of the nominally ``metallic'' 2D hole 
regime explored in Ref. \onlinecite{two} go down to as low as $r_s=26$ 
(corresponding to the lowest hole density $n=3.8 \times 10^9 cm^{-2}$ 
studied in Ref. \onlinecite{two}) with no sign of an MIT whereas the 
other systems studied in the literature exhibit the 2D MIT transition 
\cite{one} at critical $r_s$ values as low as $r_s \sim 8-12$ (Si MOSFETs) 
and $10-20$ (GaAs hole systems). The experimental results presented in 
Ref. \onlinecite{two} thus compellingly demonstrate that interaction 
(i.e., the $r_s$ parameter) is by no means the only (or perhaps even 
the dominant) variable controlling the physics of 2D MIT --- disorder 
(and perhaps even temperature) also plays an important role.

Our transport theory for the 2D hole system employs the finite temperature 
Boltzmann equation technique, which has earlier been successful in n-Si 
MOSFETs \cite{four} and n-GaAs systems \cite{five,six}. We {\it include} 
the following effects in our calculation : (1) Subband confinement 
effects (i.e., we take into account the extent of the 2D system in the 
third dimension and do not assume it  to be a zero-width 2D layer); 
(2) scattering by screened charged random impurity centers; (3) finite 
temperature and finite wave vector screening through random phase 
approximation (RPA) (actually we employ a slightly modified version 
\cite{seven} of RPA, the so-called 2D Hubbard approximation, which 
approximately and rather crudely incorporates the electron-electron 
interaction-induced vertex correction in the screening function which 
may be important at the low carrier densities \cite{two} being 
investigated --- it turns out that our calculated resistance with the 
Hubbard approximation is within $30 \%$ of the corresponding RPA results); 
(4) phonon scattering \cite{six}. 
The effects we {\it neglect} in our theory are (1) 
all localization and multiple scattering corrections; 
(2) inelastic electron interaction 
effects --- in fact, all effects of electron-electron interaction are 
neglected in our theory except for the long range screening through RPA 
and (approximate) short-range vertex correction through Hubbard approximation.

Our calculations are similar to the ones \cite{eight} we recently carried 
out for electron inversion layers in n-Si MOSFETs with two important 
differences; (1) we include the {\it full} hole density in the current 
calculations without subtracting out any critical density as done in 
Ref. \onlinecite{eight} --- this is, in fact, consistent with our Si MOSFET 
calculations since the critical density in Ref. \onlinecite{two} must be 
extremely small, and in any case SdH measurements carried out in Ref. [2]
show that {\it all} the carriers are ``free'' and participating in the 
conduction process; 
(2) we include phonon scattering effects in the current 
calculations because phonon scattering is significant for GaAs holes 
already in the $T=1-10K$ temperature range whereas phonon scattering is 
negligibly small in n-Si MOSFETs in the $1-10 K$ temperature range.
Details of phonon scattering calculations are given in Ref. \onlinecite{six}
--- the essential point is that the phonon resistivity is proportional 
to $T$ for $T > 1K$ and is negligibly small in the low temperature 
Bloch-Gr\"{u}neisen regime.

Our calculated resistivity for 2D holes in GaAs structures is shown in 
Figs. 1 and 2 for two different types of 2D quantum confinement: Square 
well (Fig. 1) and heterojunction (inversion layer type approximately 
``triangular'') confinement (Fig. 2). The qualitative results for the 
two kinds of confinement are, as expected, very similar (although the 
actual quantitative resistance values depend on the nature of confinement 
since the scattering and screening matrix elements are strongly 
confinement dependent \cite{four} through the wavefunction spread normal 
to the 2D confinement plane). The resistivity can be written as 
$\rho(T) = \rho_0 + \rho_{imp}(T) + \rho_{ph}(T)$,
where $\rho_0 \equiv \rho(T\rightarrow 0)$ is the residual resistivity 
arising entirely from (screened) charged impurity scattering in our 
theory (for a weakly localized system $\rho_0$ diverges logarithmically 
as $T \rightarrow 0$, our theory is valid above the crossover temperature 
scale for weak localization to set in --- no indication for the expected 
$\ln T $ weak localization divergence is 
seen in the experimental data of Ref. [2]
down to the lowest reported 
measurement temperature, $35 mK$).  $\rho_{ph}(T)$ is the 
resistivity contribution by phonon scattering \cite{six}
which could be quite 
significant for 2D holes in GaAs already in the $1-10K$ temperature range. 
Finally, $\rho_{imp}(T)$ is the {\it temperature dependent part} of the 
charged impurity (i.e., random disorder) scattering contribution to the 
resistivity, i.e., $\rho_0 + \rho_{imp}(T) \equiv \rho_i$ is the total 
impurity contribution to the resistivity. We note that $\rho_0$, which 
sets the overall resistivity scale [by definition, both $\rho_{imp}(T)$ 
and $\rho_{ph}(T)$ vanish as $T\rightarrow 0$] in the problem, is 
determined by the amount of the random disorder in the system which is 
in general unknown. The amount of random disorder (and consequently 
$\rho_0$) depends on the strength and the spatial distribution of all the 
impurity scattering centers in the system. We parameterize the charged 
impurity density, assuming them to be randomly distributed static Coulomb 
charged centers interacting with the 2D carriers via the screened Coulomb 
interaction. We adjust the charged impurity density (assumed to be 
randomly
distributed in our calculations) to get agreement between theory and the 
experimental data --- thus the scale $\rho_0$ is essentially an 
adjustable parameter in our theory since the actual impurity distribution 
in the 2D systems of interest is simply not known. We emphasize, however, 
that the charged impurity density needed in our theory to

\begin{figure}
\epsfysize=2.7in
\epsffile{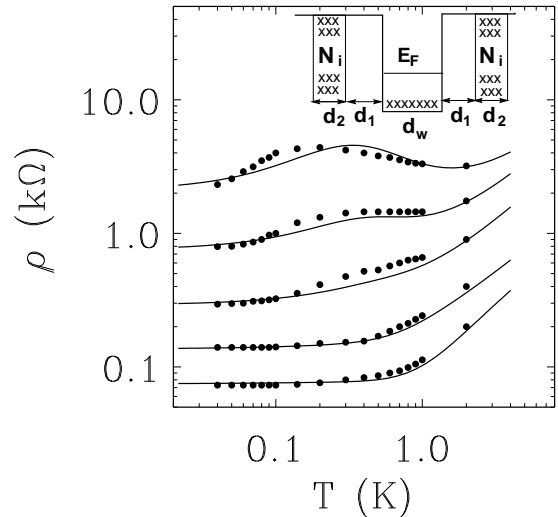}
\caption{The calculated 2D hole resistivity $\rho(T)$ for symmetric square 
well systems corresponding to the hole densities 
$n=$ 0.65, 1.07, 1.63, 3.26, 4.80$\times 10^{10}cm^{-2}$ (from top to bottom)
with random impurity densities $N_i$ = 0.7, 0.75, 0.8, 2.0, 
3.5$\times 10^{15}cm^{-3}$, respectively. 
In inset the sample 
configuration is shown schematically. In this calculation 
we use the parameters, $d_w = 300 \AA$,
$d_1 = 200 \AA$, and $d_2=50 \AA$. We use a very small random impurity 
density, $N_w = 3\times 10^{12}cm^{-3}$, in the GaAs layer itself which 
is consistent with 
the extreme high quality of the samples in Ref. [2]. Some representative 
experimental data points from Ref. [2] are shown (the actual random 
disorder in the experimental samples is unknown).
}
\end{figure}

\noindent
obtain agreement 
between our calculations and the experimental data for $\rho_0$ are 
reasonable.

Before discussing our results we make three salient remarks about our 
calculation and model. First, we neglect scattering by interface 
roughness, alloy disorder, etc. in our calculation (including only 
charged impurity scattering in the theory) since it is well-known that 
the dominant low temperature resistive mechanism in high quality GaAs 
structures arises essentially from charged impurity scattering (it is 
straightforward to include additional scattering mechanisms in our 
calculations with the unpleasant complication of having additional 
unknown parameters, such as the interface roughness strength, in the 
theory --- our choice is to keep the number of unknown adjustable 
parameters at a minimum by assuming  that all of the random disorder 
scattering is caused by randomly distributed charged impurity scattering 
which should be an excellent approximation for the extreme high quality 
GaAs samples used in Ref. [2]). Second, the Matthiessen's rule, which is 
implicitly assumed in separating out $\rho_i(T)$ and $\rho_{ph}(T)$,
is known to be not strictly valid 
at finite temperatures \cite{four} because different scattering rates do 
not simply add in the total resistivity. It is important to emphasize, 
however, that we do not assume the Matthiessen's rule in our theoretical 
calculations, and Eq. (1) is written down simply as a rough guide for 
qualitative discussion. In any case, the deviation from Matthiessen's 
rule is of the order of $30\%$ or less, which is not of much consequence 
for our discussion. Finally, the third remark we make is regarding our 
use of the single scattering Born approximation in our Boltzmann theory 
(neglecting all multiple scattering effects), which can be justified by 
noting that our calculated resistivity (and the corresponding 
experimental resistivity measured in Ref. [2]) always satisfies the 
weak scattering condition of $k_F l \gg 1$ --- in fact, our results are 
restricted to $k_F l > 3$ even in the worst situation (for our highest 
resistance results). We therefore believe that the Born approximation 
may not be a poor approximation for our problem.

In Fig. 1 we show our calculated 2D hole resistivity for symmetric square 
well systems corresponding to the sample of Ref. [2]. The actual sample 
configuration is shown schematically as an inset in Fig. 1. We also show 
some representative experimental \cite{two} results (from Fig. 2 of ref. 
[2]). We emphasize that the quantitative agreement with the data of ref. 
[2], while being certainly indicative of the essential validity of our 
theoretical approach, should not be taken too seriously --- it is 
certainly not the feature of our theory we would focus on, particularly
since the random impurity distribution in the experimental samples is 
unknown. It is the 
overall striking qualitative similarity between our microscopic theory 
and the experimental data \cite{two} which deserves attention. This is 
particularly so because the density and temperature dependence of the 
measured resistance in ref. [2] shows a throughly nontrivial non-monotonic 
behavior which is completely reproduced in our calculations. This striking 
non-monotonicity in $\rho(T)$, at lower carrier densities, arises from a 
competition among three mechanisms: Screening, which is particularly 
important at lower T; nondegeneracy and the associated quantum-classical 
crossover for $T \ge T_F$ ($\equiv E_F/k_B$, the Fermi temperature) which 
was discussed in ref. \onlinecite{eight} in the context of n-Si MOSFETs; 
and phonon scattering effect which is negligible below $1K$, but starts 
becoming quantitatively increasingly important for $T>1K$. 
The Fermi temperature for the 2D hole system can 
be expressed as $T_F = 0.64 (n/10^{10}) K$ where $n$ is the 2D hole 
density measured in units of $10^{10} cm^{-2}$. Thus for $n=4.8 \times 
10^{10}cm^{-2}$ between $n=0.65 \times 10^{10}cm^{-2}$ in Fig. 1 $T_F$ 
varies between $3K$ and $0.4K$. This makes the quantum-classical 
crossover physics particularly significant for the results of Ref. [2]
as was already noted by the authors in Ref. [2].

At higher densities (the bottom two curves in Fig. 1) the quantum-classical 
crossover effects are not particularly important because phonon scattering 
becomes important before the classical behavior \cite{eight} $\rho \sim 
T^{-1}$ can show up, and the system makes a transition from the quantum 
regime to the phonon scattering dominated regime --- the fast rise in 
$\rho(T)$ at high $T$ in Fig. 1 is the phonon scattering effect. At low 
enough densities, however, phonon scattering effects are absent (because 
phonons are frozen out in the low temperature Bloch-Gr\"{u}neisen range 
\cite{six}) {\it at
the quantum-classical crossover point} which occurs 
at very low temperatures around $T < T_F < 1K$ (the top 

\begin{figure}
\epsfysize=4.7in
\epsffile{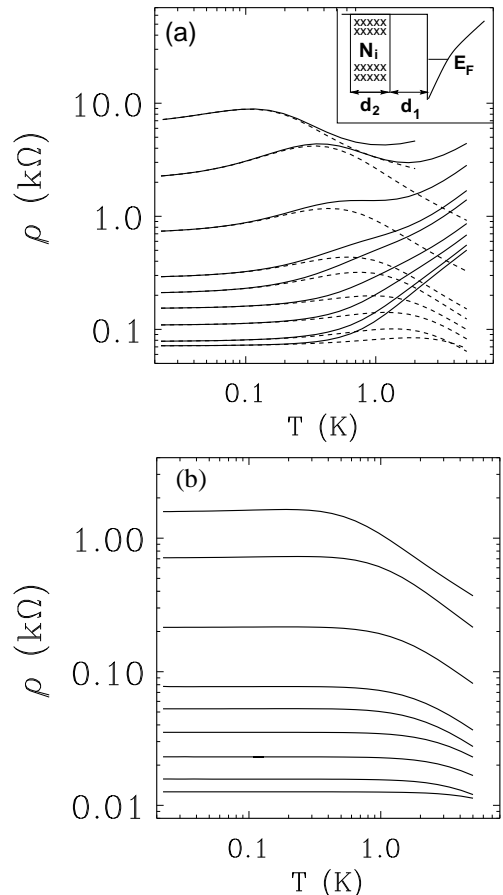}
\caption{The calculated resistivity for a heterostructure 
inversion-layer-type ``triangular'' confinement 2D (a) hole gas and 
(b) electron gas for carrier densities 
$n=$ 0.38,  0.65, 1.07, 1.63, 1.93,
3.26, 4.15, 4.80, 8.66 $\times 10^{10}cm^{-2}$ (from top to bottom) with
random impurity densities $N_i$ = 0.3, 0.3, 0.3, 
0.3, 0.3, 0.9, 1.1, 1.1, 5.0 $\times 10^{15} cm^{-3}$, respectively. 
The inset in (a) show the sample configuration. We use $d_1=250 \AA$ and
$d_2 = 100 \AA$. The impurity density in GaAs is 
$N_w = 3\times 10^{12}cm^{-3}$.
}
\end{figure}

\noindent
two curves in Fig. 1). In these low density results 
one can see $\rho(T)$ increasing with T at lower temperatures due to 
screening effects \cite{eight}, then the quantum-classical crossover 
occurs at the intermediate temperature regime around $T_F$ where 
nondegeneracy effects make resistivity decrease \cite{eight} as 
$\rho \sim T^{-1}$; eventually at higher temperatures ($ T \ge 1K$) 
phonon scattering takes over and $\rho(T)$ increases with T again. 
At higher densities $T_F$ is pushed up to the phonon scattering regime, 
and the quantum-classical crossover physics is pre-empted by phonons so 
that non-monotonicity effects are not manifest.

The non-monotonic behavior of $\rho(T)$ as a function of $n$ and $T$ is 
made more explicit in Fig. 2(a) where we show our calculated resistivity 
for the same density and temperature range as in Fig. 1 for a heterostructure 
inversion-layer-type ``triangular'' confinement 2D hole gas, separating 
out the pure impurity scattering contribution (i.e., the dashed curves 
in Fig. 2(a) leave out the phonon scattering contribution completely). First, 
we note that the resistivity results in Fig. 2(a) are very similar to those 
in Fig. 1, indicating that the transport behavior seen in ref. [2] is the 
generic behavior of a low density 2D GaAs hole system, and does not arise 
from any particular feature of the square well samples used in ref. [2]. 
Second, the interplay of screening (low temperature), phonons (high 
temperature), and nondegeneracy (high temperature and low density) is 
manifestly obvious in Fig. 2(a): the intriguing low density non-monotonicity 
in the observed $\rho(T)$ clearly arises from the fact that both screening 
and phonon scattering mechanisms give rise to a $\rho(T)$ monotonically 
increasing with $T$ (at low temperature for screening, and at high 
temperatures for phonons), but nondegeneracy effects produce a $\rho(T)$ 
decreasing with $T$ for $T\ge T_F$. Since phonon scattering is the 
dominant temperature dependent scattering mechanism in GaAs holes for 
$T>1K$, the non-monotonicity can show up in any significant way only if 
$T_F \le 1K$, which is precisely the experimental observation.

As an interesting comparison we show in Fig. 2(b) the calculated $\rho(T)$, 
without any phonon scattering, for the same densities (and impurity 
scattering parameters) as in Fig. 2(a) for a 2D {\it electron} inversion 
layers confined in a GaAs heterostructure (i.e., the only difference 
between the results for Fig.2(a) and Fig. 2(b) is that the GaAs electron 
mass has been used in the calculations corresponding to Fig. 2(b) 
rather that the hole mass.  The neglect of phonon 
scattering is justified by the fact that phonons contribute \cite{six} 
significantly to GaAs 2D electron resistivity only for $T > 10 K$ --- 
in fact, inclusion \cite{six} 
of appropriate phonon scattering would produce results indistinguishable 
from the results shown in Fig. 2(b) (i.e. upto $5K$). The difference between 
the results of Figs. 2(a) (holes) and 2(b)(electrons) is striking: there is 
essentially no observable (on log scale) temperature dependence at low 
temperatures in the 2D electron resistivity in GaAs heterostructure down 
to 2D densities as low as $n=0.38 \times 10^{10} cm^{-2}$. This essential 
temperature independence of low temperature electronic resistance in high 
quality GaAs heterostructures, which is a well-known \cite{nine} 
experimental fact, arises from the weak screening property (associated 
with its low effective mass and the associated small electronic density 
of states) of 2D electrons in GaAs heterostructures compared with 
higher mass 2D holes 
in GaAS or 2D electrons in Si MOSFETs. This weak screening behavior of 
GaAs electrons precludes any strong temperature dependent $\rho(T)$ even 
at very low carrier densities (and temperatures). The quantum-classical 
crossover phenomenon, however, still occurs around $T\sim T_F$, leading 
to a $\rho(T) \sim T^{-1}$ for $T \ge T_F$, which is manifestly obvious 
in Fig. 2(b), particularly for lower densities. Note that the Fermi 
temperature in Fig. 2(b) corresponds to $T_F = 4.1 (n/10^{10}) K$ with $n$ 
being the 2D electron density in Fig. 2(b) measured in units of 
$10^{10}cm^{-2}$. Thus the Fermi temperature in Fig. 2(b) ranges from 
$1.5K$ (top curve) to $35.5 K$ (bottom curve). 
We note that the decreasing $\rho(T)$ 
at higher $T$ in Fig. 2(b) arises not only from a quantum to classical 
crossover (which is the dominant effect at lower densities when $T_F$ 
is low), but also from the finite temperature
Fermi surface averaging in a degenerate quantum 
system. It is easy to show that the Fermi surface averaging effect at 
finite temperatures, by itself, 
always leads to a finite temperature resistivity 
which decreases weakly with temperature (even in the $T \rightarrow 0$ 
limit) -- in fact, this effect by itself leads to $\rho(T) \approx 
\rho_0[1-O(T/T_F)^2]$, and can only be observed if the temperature 
dependent screening effects are unimportant. This effect was first 
observed in 2D electrons in GaAS heterostructures more than fifteen 
years ago \cite{ten}.

To  conclude, we have developed a theory for the low temperature transport 
properties of 2D holes and electrons confined in low density and high 
mobility GaAs heterostructures. Our theory includes temperature dependent 
screening of impurity scattering and phonon scattering effects. 
Agreement between our theory and experiment suggests that screening and
impurity scattering effects play an essential 
role in determining much of the intriguing temperature and density 
dependent transport properties in 2D systems, and that random disorder 
(mostly arising from charged impurity scattering) is an important 
ingredient in the physics of low density 2D systems.

This work is supported by the U.S.-ARO. and the U.S.-ONR.


\begin{thebibliography}{99}


\bibitem{one} E. Abrahams, Physica E {\bf 3}, 69 (1998); Ann. Phys. 
(Leipzig) {\bf 8}, 539 (1999); and references therein.

\bibitem{two} A. P. Mills, Jr., A. P. Ramirez, L. N. Pfeiffer, and K. W. 
West, Phys. Rev. Lett. {\bf 83}, 2805 (1999).

\bibitem{three}E. Abrahams, P. W. Anderson, D. C. Licciardello, and T. V. 
Ramakrishnan, Phys. Rev. Lett. {\bf 42}, 
673 (1979).

\bibitem{four} T. Ando, A. B. Fowler, and F. Stern, Rev. Mod. Phys. 
{\bf 54}, 437 (1982); F. Stern, Phys. Rev. Lett. {\bf 44}, 1469 (1980);
F. Stern and S. Das Sarma, Solid State Electron. {\bf 28}, 211 (1985).

\bibitem{five}F. Stern, Appl. Phys. Lett. {\bf 43}, 974 (1983); 
K. Hirakawa and H. Sakaki, Phys. Rev. B {\bf 33}, 8291 (1986).

\bibitem{six} T. Kawamura and S. Das Sarma, Phys. Rev. B {\bf 42}, 
3725 (1990); {\it ibid} {\bf 45}, 3612 (1992); 
W. Walukiewicz, J. Appl. Phys. {\bf 59}, 3577 (1986).

\bibitem{seven} M. Jonson, J. Phys. C{\bf 9}, 3055 (1976).

\bibitem{eight}S. Das Sarma and E. H. Hwang, Phys. Rev. Lett. 
{\bf 83}, 164 (1999)

\bibitem{nine}H. L. St\"{o}rmer, L. N. Pfeiffer, K. W. Baldwin, 
and K. W. West, Phys. Rev. B {\bf 41}, 1278 (1990). See also F. W. van 
Keuls {\it et al.}, Phys. Rev. B {\bf 56}, 13263(1997).

\bibitem{ten}M. A. Paalanen, D. C. Tsui, A. C. Gossard, and J. C. M. Hwang,
Phys. Rev. B {\bf 29}, 6003 (1984).

\end{thebibliography}
\end{document}